\magnification \magstep1
\raggedbottom
\openup 2\jot 
\voffset6truemm
\centerline {\bf ON THE RIEMANN FUNCTION AND IRREGULAR SINGULAR}
\centerline {\bf POINTS FOR AXISYMMETRIC BLACK HOLE COLLISIONS}
\centerline {\bf AT THE SPEED OF LIGHT}
\vskip 1cm
\leftline {Giampiero Esposito and Cosimo Stornaiolo}
\vskip 1cm
\noindent
{\it INFN, Sezione di Napoli, Complesso Universitario di Monte
S. Angelo, Via Cintia, Edificio N', 80126 Napoli, Italy}
\vskip 0.3cm
\noindent
{\it Dipartimento di Scienze Fisiche, Universit\`a di Napoli 
Federico II, Via Cintia, Edificio N', 80126 Napoli, Italy}
\vskip 1cm
\noindent
{\bf Abstract}. The news function providing some relevant
information about angular distribution
of gravitational radiation in axisymmetric black hole collisions 
at the speed of light had been evaluated in the literature by
perturbation methods, after inverting second-order hyperbolic
operators with variable coefficients in two independent variables.
More recent work has related the appropriate Green
function to the Riemann function for such a class of hyperbolic
operators in two variables. The present paper obtains an improvement 
in the evaluation of the coefficients occurring in the second-order
equation obeyed by the Riemann function, which might prove useful 
for numerical purposes. Eventually, we find under which conditions
the original Green-function calculation reduces to finding 
solutions of an inhomogeneous second-order ordinary differential
equation with a non-regular singular point.
\vskip 100cm
The non-linear Einstein equations [1] ruling the gravitational
field are so complicated that no exact solution has been obtained
so far without making a number of simplifying assumptions. On the
other hand, in the physically more relevant case of isolated 
gravitating systems, which are time-dependent, no simplifying
assumption can be made apart from axisymmetry. Within this
framework, many efforts have been devoted to analytic and numerical
investigations of gravitational radiation produced in axisymmetric
black hole collisions at the speed of light [2-5], since such
events (although unlikely) are expected to lead to the largest 
amount of gravitational radiation ever studied at theoretical
level. 

The angular distribution of gravitational radiation is described in
part by the news function [6], and the work by D'Eath and Payne [2-5]
has shown that, in the above events, such a function can be obtained 
at second order in perturbation theory provided that one is able 
to solve a set of inhomogeneous hyperbolic equations taking
eventually the form ($m$ and $n$ being integers)
$$
{\cal L}_{m,n}\chi(q,r)=H(q,r).
\eqno (1)
$$
With the notation in Refs. [2--5], $q$ and $r$ are the independent
variables, the function $H$ is related to the source term in
the original set of equations, and the operator ${\cal L}_{m,n}$
reads
$$ \eqalignno{
{\cal L}_{m,n}&=-(2\sqrt{2}+32q){\partial^{2}\over \partial q
\partial r}+4q^{2}{\partial^{2}\over \partial q^{2}}
+64 {\partial^{2}\over \partial r^{2}}
+4(n+1)q{\partial \over \partial q} \cr
&-16n {\partial \over \partial r}+n^{2}-m^{2}.
&(2)\cr}
$$
The inverse of the operator ${\cal L}_{m,n}$ is an integral
operator whose kernel is equal to the Green function 
$G_{m,n}(q,r;q_{0},r_{0})$ of Eq. (1).

The work in Ref. [7] has however pointed out that, after reduction
of Eq. (1) to canonical form through the introduction of suitable
new variables:
$$
t \equiv \sqrt{1+16q \sqrt{2}},
\eqno (3)
$$
$$
x \equiv r+\log \left({{t-1}\over 2}\right)
-{8 \over (t-1)}-4 ,
\eqno (4)
$$
$$
y \equiv r+\log \left({{t+1}\over 2}\right)
+{8 \over (t+1)}-4 ,
\eqno (5)
$$
Eq. (1) can be solved for $\chi$ with the help of a standard
integral formula [8] which involves the Riemann function. This
is a valuable tool in the theory of hyperbolic equations in two
variables with variable coefficients, but unfortunately it has
not been much exploited (to our knowledge) in the literature on
gravitational physics. Following Ref. [7], we consider the operator
$$
L \equiv {\partial^{2}\over \partial x \partial y}
+a(x,y){\partial \over \partial x}
+b(x,y){\partial \over \partial y}+c(x,y),
\eqno (6)
$$
having defined $w$ as the function of $x-y$ such that
$$
w e^{{w^{2}-1}\over 2w}=e^{{x-y}\over 8},
\eqno (7)
$$
from which
$$
t={1+w \over 1-w},
\eqno (8)
$$
$$
a(t)=a[t(w(x-y))]={1\over 16}{(1-t)(t+1)^{2}(2n+1+t)\over
(t^{4}+4t^{2}-1)},
\eqno (9)
$$
$$
b(t)=b[t(w(x-y))]={1\over 16}{(t+1)(t-1)^{2}(2n+1-t)\over
(t^{4}+4t^{2}-1)},
\eqno (10)
$$
$$
c(t)=c[t(w(x-y))]={(m^{2}-n^{2})\over 256}
{(t-1)^{2}(t+1)^{2}\over (t^{4}+4t^{2}-1)}.
\eqno (11)
$$
The coefficient of the highest-order derivative in Eq. (6)
is constant because
$$
{\cal L}_{m,n}\chi=fL\chi=H,
$$
where $f$ is a function evaluated in Ref. [7], and hence
we study the equation
$$
L \chi={H\over f}
$$
eventually, whenever $f$ does not vanish.
The (formal) adjoint of the operator $L$ is then equal to
$$
L^{\dagger} \equiv {\partial^{2}\over \partial x \partial y}
+A(x,y){\partial \over \partial x}
+B(x,y) {\partial \over \partial y}
+C(x,y),
\eqno (12)
$$
where
$$
A \equiv -a,
\eqno (13)
$$
$$
B \equiv -b,
\eqno (14)
$$
$$
C \equiv c -{\partial a \over \partial x}
-{\partial b \over \partial y}.
\eqno (15)
$$
Let now $(\xi,\eta)$ be coordinates of a point $P$ such that
characteristics through it intersect a curve $\gamma$ at points
$A$ and $B$, $AP$ being a segment with constant $y$, and $BP$
being a segment with constant $x$. As a function of $x$ and $y$,
the Riemann function $R$ satisfies the homogeneous equation
$$
L^{\dagger}R=0,
\eqno (16)
$$
the boundary conditions 
$$
{\partial R \over \partial x}=bR \; {\rm on} \; AP,
\; \; \;
{\partial R \over \partial y}=aR \; {\rm on} \; BP,
\eqno (17)
$$
and is equal to $1$ at $P$. It is then possible to express
the solution $\chi$ of Eq. (1) in the form [7]
$$ \eqalignno{
\; & \chi(P)={1\over 2}[\chi(A)R(A)+\chi(B)R(B)]
+\int_{AB}\biggr(\left[{R\over 2}{\partial \chi \over \partial x}
+\left(bR-{1\over 2}{\partial R \over \partial x}
\right)\chi\right]dx \cr
&-\left[{R\over 2}{\partial \chi \over \partial y}
+\left(aR-{1\over 2}{\partial R \over \partial y}\right)
\chi \right]dy \biggr) \cr
&+\int \int_{\Omega}R(x,y;\xi,\eta){-H(x,y)\over 256}
{(t-1)^{2}(t+1)^{2}\over (t^{4}+4t^{2}-1)}(x,y)dx \; dy,
&(18)\cr}
$$
$\Omega$ being a suitable bounded domain. 

The main technical difficulty in Eq. (16) results from the
derivatives occurring in the coefficient $C$; since $w(x-y)$ is
found only implicitly from Eq. (7), and hence $t,a,b,c$ from
Eqs. (8)--(11), it would be helpful not having to take numerical
derivatives in (15) of a function which is only found numerically
itself. The present note solves this specific problem by pointing
out that, in Eq. (15), one has
$$
{\partial a \over \partial x}={\partial a \over \partial t}
{\partial t \over \partial w} {\partial w \over \partial x},
\eqno (19)
$$
where, from Eq. (9),
$$ \eqalignno{
\; & {\partial a \over \partial t}=-{1\over 16}\left[
{(4t^{3}+6(n+1)t^{2}+4nt-2(n+1))\over (t^{4}+4t^{2}-1)} 
\right . \cr
& \left . +{4t(1-t)(t+1)^{2}(t^{2}+2)(2n+1+t)\over
(t^{4}+4t^{2}-1)^{2}}\right],
&(20)\cr}
$$
while Eq. (8) yields
$$
{\partial t \over \partial w}={2\over (1-w)^{2}}.
\eqno (21)
$$
Moreover, from the logarithm of both sides of Eq. (7) we find
$$
\log(w)+{w\over 2}-{1\over 2w}={(x-y)\over 8}.
\eqno (22)
$$
The derivative with respect to $x$ of both sides of Eq. (22)
yields eventually
$$
{\partial w \over \partial x}={1\over 4}
{\left(1+{1\over w}\right)}^{-2}.
\eqno (23)
$$
Equations (19)--(23) yield a complete algorithm for the evaluation
of the first derivative in $C$, i.e.
$$
{\partial a \over \partial x}={1\over 2}{\partial a \over \partial t}
{w^{2}\over (1-w^{2})^{2}},
\eqno (24)
$$
where Eqs. (7), (8) and (20) should be inserted for the purpose
of numerical analysis. An entirely analogous procedure holds
for ${\partial b \over \partial y}$ in Eq. (15), i.e.
$$
{\partial b \over \partial y}={\partial b \over \partial t}
{\partial t \over \partial w}{\partial w \over \partial y},
\eqno (25)
$$
bearing in mind Eq. (10), and that
$$
{\partial w \over \partial y}=-{\partial w \over \partial x},
\eqno (26)
$$
which leads to
$$
{\partial b \over \partial y}=-{1\over 2}{\partial b \over \partial t}
{w^{2}\over (1-w^{2})^{2}},
\eqno (27)
$$
with 
$$ \eqalignno{
\; & {\partial b \over \partial t}=-{1\over 16}\left[
{(4t^{3}-6(n+1)t^{2}+4nt+2(n+1))\over (t^{4}+4t^{2}-1)} 
\right . \cr
& \left . +{4t(t+1)(t-1)^{2}(t^{2}+2)(2n+1-t)\over
(t^{4}+4t^{2}-1)^{2}}\right].
&(28)\cr}
$$
Eventually, the coefficients $A,B$ and $C$ in the operator
$L^{\dagger}$ are all expressed as functions of $w=w(x-y)$
with the help of previous equations as follows:
$$
A={w(1+n-nw)\over 4(1+2w-2w^{2}+2w^{3}+w^{4})},
\eqno (29)
$$
$$
B={w^{2}(n(-1+w)+w)\over 4(1+2w-2w^{2}+2w^{3}+w^{4})},
\eqno (30)
$$
$$ \eqalignno{ 
\; & 64(1+3w+3w^{4}+w^{5})^{2}C 
=w^{2}\Bigr[(m^{2}-n^{2})(1+4w+3w^{2}+3w^{4}+4w^{5}+w^{6}) \cr
&+4(n+1)(1-w^{2}-8w^{3}-w^{4}+w^{6})\Bigr],
&(31)\cr}
$$
where we have used Eqs. (8)--(11), (13)--(15), (20), (24),
(27) and (28). We have solved numerically Eq. (7) after
re-expressing it in the form
$$
8 \log(w)+4 \left(w -{1\over w}\right)=x-y,
\eqno (32)
$$
for $x-y$ in the closed interval $[0.8,10]$, at steps of $0.1$.
The variable $w$ is then monotonically increasing and ranges
from $1.05$ through $1.83$. The corresponding values of 
$A,B$ and $C$ have been obtained upon insertion of $w$ into
Eqs. (29)--(31). For example, at the initial point $x-y=0.8$,
the coefficients $A,B$ and $C$ read
$$
A=0.059-0.0030n,
\eqno (33)
$$
$$
B=0.065+0.0031n,
\eqno (34)
$$
$$
C=-0.0077+0.0038(m^{2}-n^{2})-0.0077n, 
\eqno (35)
$$
respectively.

Work is now in progress in applying such formulae to the numerical
evaluation of the Riemann function itself [9]. Hopefully, this will lead
to more powerful tools for the investigation of gravitational
radiation in the few cases where it is expected to be very rich,
i.e. axisymmetric black-hole collisions at the speed of light [2--5].

Meanwhile, we are also making progress on alternative approaches to
finding solutions of the equation $L \chi={H \over f}={\widetilde H}$
that we started with. To exploit both Fourier transform theory and
the fact that the coefficients in the operator $L$ depend on $t$ and
hence, eventually, only on the difference $x-y$, we give up the
canonical form of the operator $L$ by introducing the independent
variables
$$
X \equiv {x+y \over 2}, \; \; \; 
Y \equiv {x-y \over 2}.
\eqno (36)
$$
Our inhomogeneous hyperbolic equation is then turned into
$$
{\widetilde L}\chi(X,Y)=4{\widetilde H}(X,Y) \equiv h(X,Y),
\eqno (37)
$$
where
$$
{\widetilde L} \equiv {\partial^{2}\over \partial X^{2}}
-{\partial^{2}\over \partial Y^{2}}
+2(A+B){\partial \over \partial X}
+2(A-B){\partial \over \partial Y}+4C,
\eqno (38)
$$
with coefficients depending only on $Y$. It is therefore more
convenient to consider the Fourier transform of $\chi$ with
respect to $X$ (assuming that it exists), i.e.
$$
{\widehat \chi}(k,Y) \equiv {1\over \sqrt{2\pi}}
\int_{-\infty}^{\infty} \chi(X,Y)e^{-ikX}dX,
\eqno (39)
$$
with inversion formula
$$
\chi(X,Y)={1\over \sqrt{2\pi}}
\int_{-\infty}^{\infty} {\widehat \chi}(k,X)
e^{ikX}dk.
\eqno (40)
$$
If also the right-hand side of Eq. (37) admits Fourier transform
${\widehat h}(k,Y)$ with respect to $X$, we therefore obtain the
second-order equation
$$
\left[{\partial^{2}\over \partial Y^{2}}
+2(B-A){\partial \over \partial Y}
+(k^{2}-4C-2ik(A+B))\right]{\widehat \chi}(k,Y)
=-{\widehat h}(k,Y).
\eqno (41)
$$
By virtue of Eqs. (9), (10), (13) and (14) we find
$$
B-A={1\over 8}{(n+2)t(1-t^{2})\over (t^{4}+4t^{2}-1)},
\eqno (42)
$$
$$
B+A={1\over 8}{(t^{2}-1)(t^{2}+2n+1)\over (t^{4}+4t^{2}-1)},
\eqno (43)
$$
and we re-express Eq. (32) in the form
$$
4\log(w)+2 \left(w-{1\over w}\right)=Y.
\eqno (44)
$$
Note now that, for each fixed value of $k$, Eq. (41) may be viewed
as an inhomogeneous second-order ordinary differential equation
for $\widehat \chi$. On denoting by $M$ the operator in square
brackets on the left-hand side of Eq. (41), and by $\chi_{1}$ and
$\chi_{2}$ two linearly independent integrals of the homogeneous
equation $M {\widehat \chi}=0$, one can use the method of variation
of parameters, according to which the general solution of Eq.
(41) reads
$$
{\widehat \chi}=\lambda_{1}\chi_{1}(Y)+\lambda_{2}\chi_{2}(Y)
+v_{1}(Y)\chi_{1}(Y)+v_{2}(Y)\chi_{2}(Y),
\eqno (45)
$$
with $\lambda_{1}$ and $\lambda_{2}$ constants, while
$v_{1}$ and $v_{2}$ are chosen in such a way that
$$
v_{1}'\chi_{1}+v_{2}'\chi_{2}=0,
\eqno (46)
$$
$$
v_{1}'\chi_{1}'+v_{2}'\chi_{2}'=-{\widehat h} \equiv j(Y),
\eqno (47) 
$$
the prime denoting derivative with respect to $Y$. The system 
expressed by Eqs. (46) and (47) is solved by
$$
v_{1}(Y)=-\int {\chi_{2}(Y)j(Y)\over W(Y)}dY,
\eqno (48)
$$
$$
v_{2}(Y)=\int {\chi_{1}(Y)j(Y)\over W(Y)}dY,
\eqno (49)
$$
where $W$ is the standard notation for the Wronskian of $\chi_{1}$
and $\chi_{2}$, i.e.
$$
W(Y) \equiv \chi_{1}(Y)\chi_{2}'(Y)-\chi_{2}(Y)\chi_{1}'(Y).
\eqno (50)
$$
In the course of investigating the singular
points of Eq. (41), the technical difficulty results from the
fact that its coefficients are ratios of polynomials in the $t$
or $w$ variable, but not in the $Y$ variable itself by virtue 
of Eq. (44). A way out is obtained by turning
Eq. (41) into an ordinary differential equation with $w$ taken
as the independent variable. For this purpose, we exploit Eq. (44)
to find 
$$
{dw\over dY}={1\over 2}\left(1+{1\over w}\right)^{-2},
\eqno (51)
$$
and hence
$$
\left[{d^{2}\over dw^{2}}
+g_{1}(w){d \over dw}+g_{2}(w)\right]
{\widehat \chi}(k,w)=\sigma(w),
\eqno (52)
$$
where
$$ \eqalignno{
\; & g_{1}(w) \equiv {2\over w^{2}}
\left(1+{1\over w}\right)^{-1}+4(B-A)
\left(1+{1\over w}\right)^{2} \cr
&={2\over w(1+w)}-{(n+2)\over 2}
{(1+w)^{2}(1-w^{2})\over w (1+(1+\sqrt{5})w+w^{2})
(1+(1-\sqrt{5})w+w^{2})},
&(53)\cr}
$$
$$ \eqalignno{
\; & g_{2}(w) \equiv 4 \left(1+{1\over w}\right)^{4}
(k^{2}-4C-2ik(A+B)) \cr
&=4\biggr[{(1+w)^{4}\over w^{4}}k^{2}-{1\over 16w^{2}}
\left((m^{2}-n^{2})
{(1+w)^{4}\over
(1+(1+\sqrt{5})w+w^{2})(1+(1-\sqrt{5})w+w^{2})} \right . \cr
& \left . +4(n+1)
{(1-w^{2}-8w^{3}-w^{4}+w^{6})(1+w)^{2}
\over (1+(1+\sqrt{5})w+w^{2})^{2}(1+(1-\sqrt{5})w+w^{2})^{2}}\right) \cr
&-{ik\over 2}{((n+1)(1+w^{2})-2nw)(1+w)^{4}\over 
w^{3}(1+(1+\sqrt{5})w+w^{2})(1+(1-\sqrt{5})w+w^{2})}\biggr],
&(54)\cr}
$$
$$
\sigma(w) \equiv 4 \left(1+{1\over w}\right)^{4}
j(Y(w)).
\eqno (55)
$$
Equation (52) has therefore a non-regular singular point at $w=0$,
in that the coefficient $g_{2}(w)$ has a fourth-order pole
therein. This corresponds to the point $x=-\infty$ in the original
variable defined in Eq. (4). Moreover, the polynomial
$p(w) \equiv 1+(1+\sqrt{5})w+w^{2}$ has real roots equal to
$$
w_{1,2} \equiv 
-{1\over 2}(1+\sqrt{5}) \pm \sqrt{{1\over 2}(1+\sqrt{5})}.
\eqno (56)
$$
At these points, $g_{1}(w)$ has a first-order pole, and
$g_{2}(w)$ has a second-order pole. The last pole occurs for
$g_{1}(w)$ at $w=-1$, and it is of first order. No further poles
occur at finite real values of $w$, since the polynomial
$q(w) \equiv 1+(1-\sqrt{5})w+w^{2}$ has the complex conjugate roots
$$
{1\over 2}(\sqrt{5}-1)\pm i \sqrt{{1\over 2}(\sqrt{5}-1)}.
$$

By virtue of (54), Eq. (52) does not possess {\it normal integrals}.
These are meant to be integrals admitting the factorization 
$e^{\Omega}u$ [10], where $\Omega$ is a polynomial in ${1\over w}$ and
$u$ solves an equation with a Fuchsian singularity at the origin.
The point $w=0$ would be therefore an essential singularity through
the occurrence of the factor $e^{\Omega}$. In our case, on 
re-expressing $g_{1}$ and $g_{2}$ in the form
$$
g_{1}(w)={2\over w(1+w)}+{F_{1}\over w},
$$
$$
g_{2}(w)=4{(1+w)^{4}\over w^{4}}k^{2}+{F_{2}\over w^{2}}
+{F_{3}\over w^{3}},
$$
with obvious meaning of the functions $F_{1},F_{2},F_{3}$ by
comparison with (53) and (54), 
the ansatz for normal integrals yields the 
following second-order equation for $u$:
$$
\left[{d^{2}\over dw^{2}}+(2\Omega'+g_{1}){d\over dw}
+(\Omega''+{\Omega'}^{2}+g_{1}\Omega'+g_{2})\right]
u=e^{-\Omega}\sigma.
\eqno (57)
$$
Since $u$ should obey an equation with regular singular point
at the origin, we try to choose $\Omega$ in such a way that
the coefficients of $w^{-4}$ and $w^{-3}$ vanish in Eq. (56).
The former task is accomplished by choosing
$$
\Omega'=2ik\left(1+{1\over w}\right)^{2},
$$
which yields
$$ 
\Omega''+{\Omega' \over w}\left(F_{1}+{2\over (1+w)}\right)
+{F_{3}\over w^{3}} 
={1\over w^{3}}\left(F_{3}+2ik(1+w)^{2}F_{1}+4ikw \right)
+{\rm O}(w^{-2}),
\eqno (58) 
$$
where the term in round brackets multiplying $w^{-3}$ does not 
vanish. This is why normal integrals cannot be found in our problem.

We have therefore to look for ${\widehat \chi}$ in the most
general form [10]
$$
{\widehat \chi}(k,w)=w^{\gamma}\sum_{l=-\infty}^{\infty}
a_{l}w^{l}, \; \; \; w \in ]0,w_{1}[,
\eqno (59)
$$
where $\gamma$ accounts for the multi-valuedness of the solution
(any integer part being absorbed in the Laurent expansion valid
in the open interval $w \in ]0,w_{1}[$). This expansion is now
inserted into the homogeneous equation associated to Eq. (52), i.e.
$$
\left[{d^{2}\over dw^{2}}+g_{1}(w){d\over dw}+g_{2}(w)\right]
{\widehat \chi}(k,w)=0,
\eqno (60)
$$
since the knowledge of two linearly independent integrals of
Eq. (60) is sufficient to solve Eq. (52) by exploiting the method
of variation of parameters previously described. For our purposes
we consider Laurent expansions of $g_{1}$ and $g_{2}$ in the form
$$
g_{1}(w)=\sum_{s=-\infty}^{\infty}a_{s}^{0}w^{s}
={h_{1}(w)\over w}, \; \; w \in ]0,w_{1}[,
\eqno (61)
$$
$$
g_{2}(w)=\sum_{s=-\infty}^{\infty}b_{s}^{0}w^{s}
={h_{2}(w)\over w^{4}}, \; \; w \in ]0,w_{1}[.
\eqno (62)
$$
We therefore get from Eq. (60) the condition
$$
\sum_{p=-\infty}^{\infty}\left[\sum_{l=-\infty}^{\infty}
J(\gamma;l,p,k)a_{l} \right]w^{p}=0,
\eqno (63)
$$
having defined
$$
J(\gamma;l,p,k) \equiv (\gamma+l)(\gamma+l-1)\delta_{l,p+2}
+(\gamma+l)a_{p-l+1}^{0}(k)+b_{p-l}^{0}(k).
\eqno (64)
$$
Equation (63) leads to the infinite system of equations
$$
\sum_{l=-\infty}^{\infty} J(\gamma;l,p,k)a_{l}=0,
\eqno (65)
$$
for all $p=-\infty,...,+\infty$, where 
the coefficients $a_{l}^{0}$ and $b_{l}^{0}$ 
occurring in $J$ are all known. Indeed one finds
(cf. (53) and (54))
$$
g_{1}(w)={1-{n\over 2}-2w-nw^{2}+{\rm O}(w^{3}) \over w},
\eqno (66)
$$
$$
g_{2}(w)={(-1+4k^{2}-n)+2(1+8k^{2}+n-ik(1+n))w
+{\rm O}(w^{2}) \over w^{4}},
\eqno (67)
$$
if $w \in ]0,w_{1}[$. By virtue of (64), Eq. (65) may be cast
in the form
$$
a_{p}=\sum_{l=-\infty}^{\infty}G_{\gamma}(p,l)a_{l},
\eqno (68)
$$
having defined
$$
G_{\gamma}(p,l) \equiv -{1\over (\gamma+p)(\gamma+p-1)}
\Bigr[(\gamma+l)a_{p-l-1}^{0}+b_{p-l-2}^{0}\Bigr],
\eqno (69)
$$
where the general form of the coefficients $a_{l}^{0}$ and 
$b_{l}^{0}$ is found to be, in agreement with (53) and (54),
$$
a_{l}^{0} \equiv n \alpha_{1,l}+\alpha_{2,l},
\eqno (70)
$$
$$
b_{l}^{0} \equiv m^{2}\alpha_{3,l}+n^{2}\alpha_{4,l}
+n \alpha_{5,l}+\alpha_{6,l}.
\eqno (71)
$$
The numerical coefficients $\alpha_{k,l}$, for all $k=1,...,6$,
can be inferred from (66) and (67). For example, one finds
$$
\alpha_{1,1}=-{1\over 2}, \; \; \alpha_{2,1}=1, \; \; 
\alpha_{1,2}=-2, \; \; \alpha_{2,2}=-2, \; \; 
\alpha_{1,3}=7, \; \; \alpha_{2,3}=12.
$$
The polydromy parameter $\gamma$ is found, at least in principle,
by requiring that the homogeneous linear system expressed by 
Eq. (68) should have non-trivial solutions. Such a problem is 
currently under investigation. Hopefully, an intriguing link between
black hole physics and the theory of infinite determinants [10]
will be found to emerge.
\vskip 0.3cm
\leftline {\bf Acknowledgments}
\vskip 0.3cm
\noindent
Our work has been partially supported by the Progetto
di Ricerca di Interesse Nazionale {\it SINTESI 2000}.
G. Esposito is grateful to Peter D'Eath for encouragement.
\vskip 0.3cm
\leftline {\bf References}
\vskip 0.3cm
\noindent
\item {[1]}
A. Einstein, {\it Ann. der Phys.} {\bf 49}, 769 (1916).
\item {[2]}
P. D. D'Eath and P. N. Payne, {\it Phys. Rev.} {\bf D 46},
658 (1992).
\item {[3]}
P. D. D'Eath and P. N. Payne, {\it Phys. Rev.} {\bf D 46},
675 (1992).
\item {[4]}
P. D. D'Eath and P. N. Payne, {\it Phys. Rev.} {\bf D 46},
694 (1992). 
\item {[5]}
P. D. D'Eath, {\it Black Holes: Gravitational Interactions}
(Oxford, Clarendon, 1996).
\item {[6]}
H. Bondi, M. G. J. van der Berg and A. W. K. Metzner,
{\it Proc. R. Soc. London} {\bf A 269}, 21 (1962).
\item {[7]}
G. Esposito, {\it Class. Quantum Grav.} {\bf 18}, 1997 (2001).
\item {[8]}
R. Courant and D. Hilbert, {\it Methods of Mathematical Physics.
II. Partial Differential Equations} (New York, Interscience, 1961).
\item {[9]}
S. Alexeyev, G. Esposito and C. Stornaiolo (in preparation).
\item {[10]}
A. R. Forsyth, {\it Theory of Differential Equations. Part III,
Vol. IV} (New York, Dover, 1959).

\bye